\documentclass[apl,superscriptaddress,preprint,endfloats,dvipdfmx]{revtex4-1}
\newcommand {\ECS}{EuCdSb$_{\mathrm{2}}$}

\newcommand {\etal}{\textit{et al}.}
\usepackage{graphicx}
\usepackage{bm}
\usepackage{pifont}
\usepackage{amsmath}
\usepackage{amssymb}
\begin{document}
\title{
Molecular beam deposition of a new layered pnictide\\ with distorted Sb square nets
}
\author{M. Ohno}
\affiliation{Department of Applied Physics and Quantum-Phase Electronics Center (QPEC), the University of Tokyo, Tokyo 113-8656, Japan}
\author{M. Uchida}
\email[Author to whom correspondence should be addressed: ]{m.uchida@phys.titech.ac.jp}
\affiliation{Department of Applied Physics and Quantum-Phase Electronics Center (QPEC), the University of Tokyo, Tokyo 113-8656, Japan}
\affiliation{PRESTO, Japan Science and Technology Agency (JST), Tokyo 102-0076, Japan}
\affiliation{Department of Physics, Tokyo Institute of Technology, Tokyo 152-8550, Japan}
\author{Y. Nakazawa}
\affiliation{Department of Applied Physics and Quantum-Phase Electronics Center (QPEC), the University of Tokyo, Tokyo 113-8656, Japan}
\author{S. Sato}
\affiliation{Department of Applied Physics and Quantum-Phase Electronics Center (QPEC), the University of Tokyo, Tokyo 113-8656, Japan}
\author{M. Kriener}
\affiliation{RIKEN Center for Emergent Matter Science (CEMS), Wako 351-0198, Japan}
\author{A. Miyake}
\affiliation{The Institute for Solid State Physics (ISSP), the University of Tokyo, Kashiwa, Chiba 277-8581, Japan}
\author{M. Tokunaga}
\affiliation{RIKEN Center for Emergent Matter Science (CEMS), Wako 351-0198, Japan}
\affiliation{The Institute for Solid State Physics (ISSP), the University of Tokyo, Kashiwa, Chiba 277-8581, Japan}
\author{Y. Taguchi}
\affiliation{RIKEN Center for Emergent Matter Science (CEMS), Wako 351-0198, Japan}
\author{M. Kawasaki}
\affiliation{Department of Applied Physics and Quantum-Phase Electronics Center (QPEC), the University of Tokyo, Tokyo 113-8656, Japan}
\affiliation{RIKEN Center for Emergent Matter Science (CEMS), Wako 351-0198, Japan}
\begin{abstract}
While the family of layered  pnictides $ABX_2$ ($A$ : rare or alkaline earth metals, $B$ : transition metals, $X$ : Sb/Bi) can host Dirac dispersions based on Sb/Bi square nets, nearly half of them has not been synthesized yet for possible combinations of the $A$ and $B$ cations.
Here we report the fabrication of {\ECS} with the largest $B$-site ionic radius, which is stabilized for the first time in thin film form by molecular beam deposition.
{\ECS} crystallizes in an orthorhombic $Pnma$ structure and exhibits antiferromagnetic ordering of the Eu magnetic moments at $T_\mathrm{N}=15$~K.
Our successful growth will be an important step for further exploring novel Dirac materials using film techniques.
\end{abstract}
\maketitle
Dirac materials with linear energy dispersions have attracted growing attention in the light of exploring new compounds and elucidating their magnetotransport \cite{Wehling2014}.
In particular, Dirac dispersions interacting with magnetism have significant potential for producing rich quantum transport \cite{Jungwirth2018,Smejkal2018,Shao2019a}.
Recently, ternary layered pnictides $ABX_2$ ($A$ : rare or alkaline earths, $B$ : transition metals, $X$ : Sb/Bi) have been reported to host highly anisotropic Dirac or Weyl dispersions \cite{Masuda2016,Masuda2018,Liu2019_arxiv,Sakai2020a,Ling2018, Kealhofer2018a,Weber2018,Ramankutty2018a,Qiu2019,Soh2019b, Park2011,Wang2012, Lee2013,Yan2017b,Liu2017f,Klemenz2019,Klemenz2020,Takahashi2020a}, and unprecedented quantum transport has been observed originating from the interplay between the Dirac dispersion and magnetic ordering, as exemplified by bulk half-integer quantum Hall effect in EuMnBi$_2$ \cite{Masuda2016,Masuda2018}, BaMnSb$_2$ \cite{Liu2019_arxiv, Sakai2020a}, and SrMnSb$_2$ \cite{Ling2018}.
Therefore, exploring new $ABX_2$ compounds especially with magnetic $A$ and/or $B$ ions is important for systematically investigating their unique magnetotransport.

As summarized in Fig.~1(a), there are four space groups ($I$4/$mmm$, $Imm$2, $P$4/$nmm$, and $Pnma$) for $AB$Sb$_2$.
$AB$Sb$_2$ have two types of Sb atoms in the unit cell: one with $-1$ valence (defined as Sb1) consisting of the Sb net which hosts Dirac or Weyl dispersions and the other with $-3$ valence (defined as Sb2) forming Sb tetrahedra around the $B$-site cations, which are separated by $A$ layers.
The $A$-site cations are coincidently stacked across the Sb1 net in $I4/mmm$ and $Imm2$, while they are staggeredly stacked in $P4/nmm$ and $Pnma$.
The Sb1 square nets are slightly distorted in $Pnma$ and $Imm2$, resulting in a zig-zag chainlike structure along the $b$-axis.

As seen in the map in Fig.~1(b), it becomes difficult to stabilize $AB$Sb$_2$ with smaller $A$-site and larger $B$-site ionic radii, and nearly half of $AB$Sb$_2$ has not been synthesized yet for all the possible combinations of the $A$ and $B$ cations \cite{Weber2018,Ramankutty2018a,Kealhofer2018a,Qiu2019,Soh2019b,Cordier1977,Brechtel1981,Yi2017,Gong2020b,You2019,May2009,Park2016a,Liu2016g,Huang2017a,Huang2020,He2017b,Ling2018,Wang2018f,Liu2019f,Wang2020c,Liu2019e}.
In this lower right region of the map in Fig.~1(b), phase separation may occur, for example, from {\ECS} to EuSb$_2$ with $-1$ valence \cite{Hulliger1978, Ohno2020} and EuCd$_2$Sb$_2$ with $-3$ valence \cite{Su2020}, because $AB$Sb$_2$ has different $-1$ and $-3$ valence states in one structure.
Especially in the case of $B=$~Cd, its high vapor pressure also makes the fabrication difficult.

In this context, developing techniques for stabilizing new $AB$Sb$_2$ compounds is strongly called for advancing this research field.
Here we report the fabrication of its new member {\ECS}, which is stabilized for the first time in thin film form by molecular beam deposition.
{\ECS} crystallizes in the orthorhombic $Pnma$ structure and exhibits unique nonmetallic transport.

Single-crystalline (0001) Al$_2$O$_3$ substrates were annealed at 850~${^\circ}$C in a base pressure about $3\times10^{-7}$~Pa and then EuCdSb$_2$ films were grown on it in an Epiquest RC1100 chamber \cite{Nakazawa2019}.
The molecular beams were simultaneously provided from conventional Knudsen cells containing 3N Eu, 6N Cd, and 6N Sb.
The growth temperature was set at 390~${^\circ}$C, and the beam equivalent pressures, measured by an ionization gauge, were set to 1.2${\times}$10${^{-5}}$~Pa for Eu, 5.0${\times}$10${^{-4}}$~Pa for Cd, and 8.5${\times}$10${^{-6}}$~Pa for Sb (for details see Supplementary Materials \cite{Supplement}).
To avoid Cd deficiency, the Cd flux was set 40 times higher than the Eu flux.
The film thickness was typically set at 25~nm for structural characterization and magnetotransport measurements, and 50~nm for magnetization measurements.
The growth rate was about 0.07~$\mathrm{\AA}$/s.
Magnetotransport was measured by a standard four-probe method for 200~$\mu$m-width multi-terminal Hall bars.
Longitudinal and Hall resistivities were measured up to 22.4~T using a non-destructive pulsed magnet \cite{Uchida2017} at the International MegaGauss Science Laboratory in the Institute for Solid State Physics at the University of Tokyo.
Temperature dependence of the resistivity and magnetization was measured using a Quantum Design Physical Properties Measurement System and a Magnetic Property Measurement System, respectively.

Figure~2 shows structural characterization of an obtained {\ECS} film.
As shown in the x-ray diffraction (XRD) $\theta$-2$\theta$ scan in Fig.~2(a), the reflections from the (200) {\ECS} lattice planes are observed without any impurity phases.
Its out-of-plane lattice constant along the $a$-axis is calculated to be $22.53$~{\AA}.
As confirmed in Fig.~2(b), clear Laue fringes indicate highly coherent lattice ordering along the out-of-plane direction.
A rocking curve taken for the (800) film peak in Fig.~2(c) is very sharp with a full width at half maximum of $0.11$~degrees, ensuring high crystallinity of the film.
On the other hand, the in-plane XRD reciprocal space map shown in Fig.~2(d) suggests that the $b$- and $c$-axes are rather randomly oriented forming domain structures.
In-plane $(020)$ and $(002)$ EuCdSb$_2$ peaks are also indiscernible, indicating that the in-plane lattice constants of the $b$- and $c$-axes are almost the same.
The atomic force microscopy image in Fig.~2(e) reveals a flat surface with a root mean square (RMS) roughness of $0.27$~nm.

The crystal structure of this new compound is closely examined by taking high-angle annular dark-field scanning transmission electron microscopy (HAADF-STEM) images.
A cross-sectional image of the {\ECS} film in Fig.~3(a) shows clear stacking of the Sb square nets and other layers.
Its in-plane lattice constant along the $c$-axis is calculated to be 4.47~{\AA} by comparing with the $a$-axis length determined by the XRD $\theta$-2$\theta$ scan.
Higher resolution HAADF-STEM image and corresponding elemental maps in Fig.~3(b)-3(e) reveal additional details of the atomic arrangement.
Staggered stacking of Eu atoms across the Sb1 layers rules out the $I4/mmm$ and $Imm2$ structures among the four possible space groups as shown in Fig.~1(a).
As confirmed in the intensity profiles in Fig.~3(f), moreover, alternate left and right shifts of the Sb atoms are clearly resolved in the Sb1 layers, in contrast to the equally spaced Cd atoms.
These Sb shifts correspond to the distorted Sb square nets as shown in  Fig.~3(g) and thus we conclude that the {\ECS} crystal structure is $Pnma$.
The $Pnma$ structure is a nonsymmorphic structure which protects band crossings of Dirac dispersions \cite{Klemenz2020}.

Figure~4 summarizes fundamental transport and magnetic properties of the {\ECS} films.
Temperature dependence of the magnetization in Fig.~4(a) shows a kink at $T_\mathrm{N}=15$~K, which is ascribed to antiferromagnetic (AFM) ordering of the Eu magnetic moments as detailed in the following.
This kink is observed much sharper for the in-plane field, while an almost constant behavior below $T_\mathrm{N}$ is found for the out-of-plane field.
This indicates that the Eu magnetic moments are antiferromagnetically ordered in the in-plane.
Similar to other Eu$BX_2$ compounds \cite{Masuda2016,Masuda2018,Yi2017,Soh2019b,Gong2020b, Wang2020c}, the Eu magnetic moments of EuCdSb$_2$ probably forms the A-type AFM structure where ferromagnetic (FM) in-plane layers are antiferromagnetically stacked along the out-of-plane direction.
Figure~4(b) shows the nonmetallic temperature dependence of the longitudinal resistivity.
A drastic upturn at $15$~K coincides with the AFM ordering, as also confirmed by its shift upon increasing the magnetic field.
This nonmetallic temperature dependence is a common feature only with EuMnSb$_2$ \cite{Yi2017,Soh2019b,Gong2020b}, and the further upturn below $T_\mathrm{N}$ appears only in {\ECS}, while other $AB$Sb$_2$ compounds shown in Fig.~1(b) exhibit metallic behavior \cite{He2017b,May2009,Park2016a,Liu2016g,Huang2017a,Ling2018,Kealhofer2018a,Wang2018f,Liu2019f,You2019,Huang2020,Wang2020c,Liu2019e}.

Figure~4(c) presents magnetoresistance (MR) taken by out-of-plane magnetic field sweeps at various temperatures.
Negative MR at the base temperature of $1.4$~K saturates at $B_\mathrm{s}=12.7$~T, which shifts to lower fields with increasing temperature and then disappears above ${T_{\mathrm{N}}}$.
Therefore, it can be understood that this saturation corresponds to the phase transition from an AFM to a forced ferromagnetic (FM) phase and electron scattering by the Eu magnetic moments is suppressed through this canting process.
For the in-plane field as shown in the inset, on the other hand, a transition from positive to negative MR is observed.
This is ascribed to a spin-flop transition, consistent with the in-plane AFM ordering in the ground state (see Supplementary Materials \cite{Supplement}).
After the spin-flop transition, a similar negative MR is observed also for the in-plane field.

Finally, Figure~4(d) shows Hall resistivity measured at various temperatures.
While more than one types of carriers contribute in the conduction in most of the $ABX_2$, an almost linear Hall resistivity is observed for {\ECS} in the entire temperature range.
Carrier density and mobility are estimated at 1.9 $\times$ $10^{19}$ cm$^{-3}$ and $86$~cm$^{2}$/Vs by single-carrier fitting at $1.4$~K.
Despite the large Eu magnetic moments, there are no clear indications of anomalous Hall effect for {\ECS}, similar to other magnetic $AB$Sb$_2$ \cite{Park2016a,Liu2016g, He2017b,Yi2017,Huang2017a, Ling2018,Kealhofer2018a,Wang2018f,Liu2019f,Liu2019_arxiv,Sakai2020a,Wang2020c,Huang2020}.

In summary, we have demonstrated the fabrication of a new $AB$Sb$_2$ compound using molecular beam deposition.
{\ECS} crystallizes in orthorhombic $Pnma$ structure and exhibits unique nonmetallic transport properties in addition to in-plane AFM ordering.
Our successful growth of {\ECS} by film-technique paves the way for further exploring $ABX_2$ systems and designing their heterostructures.

\subsection*{SUPPLEMENTARY MATERIAL}
Phase diagram for optimizing growth condition, wide-range TEM image, and magnetization curves confirming the spin flop transition are provided in the supplementary material.

\subsection*{ACKNOWLEDGMENTS}
The authors would like to thank H. Sakai for helpful discussions.
This work was supported by JST PRESTO Grant No. JPMJPR18L2 and JST CREST Grant No. JPMJCR16F1, Japan and by Grant-in-Aid for Scientific Research (B) No. JP18H01866 from MEXT, Japan. The data that support the findings of this study are available from the corresponding author upon reasonable request.

\subsection*{DATA AVAILABILITY}
The data that support the findings of this study are available from the corresponding author upon reasonable request.

\clearpage

\begin{figure}
	\begin{center}
		\includegraphics*[bb=0 0 816 551,width=16cm]{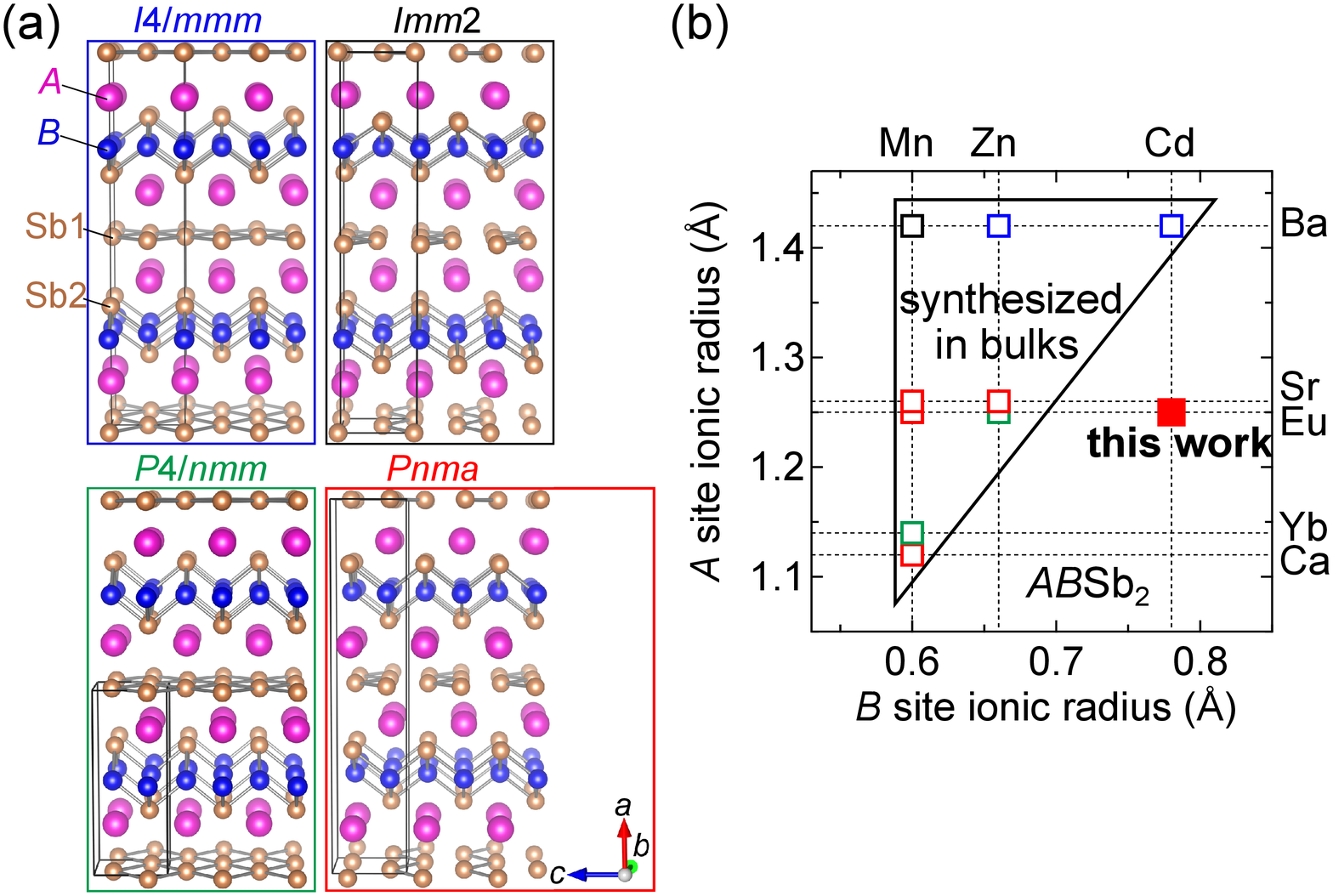}
		\caption{
			(a) Crystal structures of the three space groups in $AB$Sb$_2$ systems ($I4$/$mmm$, $Imm2$, $P4$/$nmm$, and $Pnma$).
			They have two types of Sb atoms (valence states of $-1$ for Sb1 and $-3$ for Sb2), which are separated by $A$ layers.
			One unit cell of each structure is represented by black lines.
			(b) Mapping of previously synthesized $AB$Sb$_2$ bulks (open squares) \cite{Weber2018,Ramankutty2018a,Kealhofer2018a,Qiu2019,Soh2019b,Cordier1977,Brechtel1981,Yi2017,Gong2020b,You2019,May2009,Park2016a,Liu2016g,Huang2017a,Huang2020,He2017b,Ling2018,Wang2018f,Liu2019f,Wang2020c,Liu2019e} and the newly stabilized {\ECS} film in this study (filled square), as a function of $A$ ($=$~Ca, Yb, Eu, Sr, and Ba) and $B$ ($=$~Mn, Zn, and Cd)-site ionic radii.
			Depending on their space groups, the squares are colored in blue ($I4$/$mmm$), black ($Imm2$), green ($P4$/$nmm$), or red ($Pnma$).
			This map clearly shows that the $P4/nmm$ or $Pnma$ structure becomes stable for smaller, while $I4$/$mmm$ and $Imm2$ do for larger $A$-site ionic radius.
			Regarding BaMnSb$_2$, its detailed structure has been recently found $Imm2$ \cite{Liu2019_arxiv, Sakai2020a}, while it was originally characterized as $I4/mmm$ \cite{Liu2016g, Huang2017a, Huang2020}.
		}
		\label{fig1}
	\end{center}
\end{figure}
\begin{figure}
	\begin{center}
		\includegraphics*[bb=0 0 505 428,width=14cm]{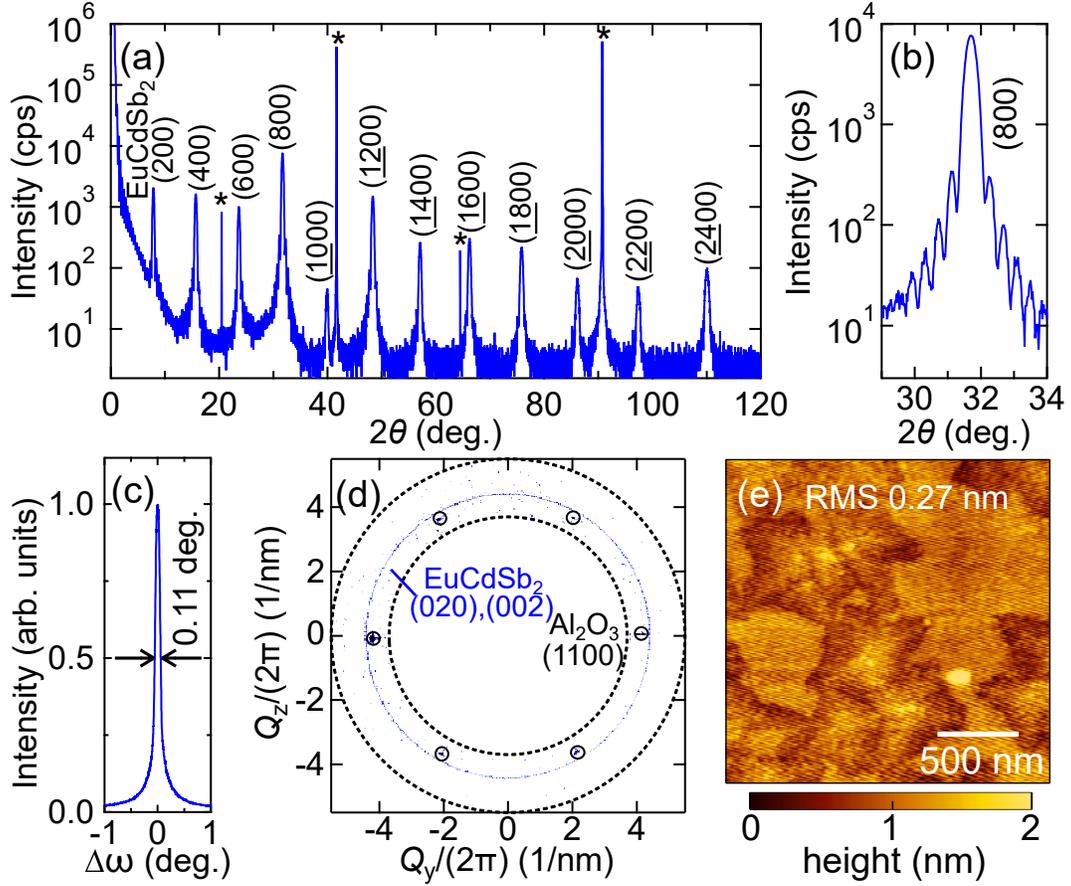}
		\caption{
			(a) XRD $\theta$-2$\theta$ scan of a {\ECS} film grown on a (0001) Al${_{2}}$O${_{3}}$ substrate.
			Al$_2$O$_3$ substrate peaks are marked with an asterisk.
			(b) Enlarged view of the $\theta$-2$\theta$ scan and (c) rocking curve around the (800) {\ECS} film peak.
			(d) In-plane reciprocal space mapping measured by grazing incidence.
			(e) Surface morphology of the film, taken by atomic force microscopy.
		}
		\label{fig2}
	\end{center}
\end{figure}
\begin{figure}
	\begin{center}
		\includegraphics*[bb=0 0 1130 589,width=16cm]{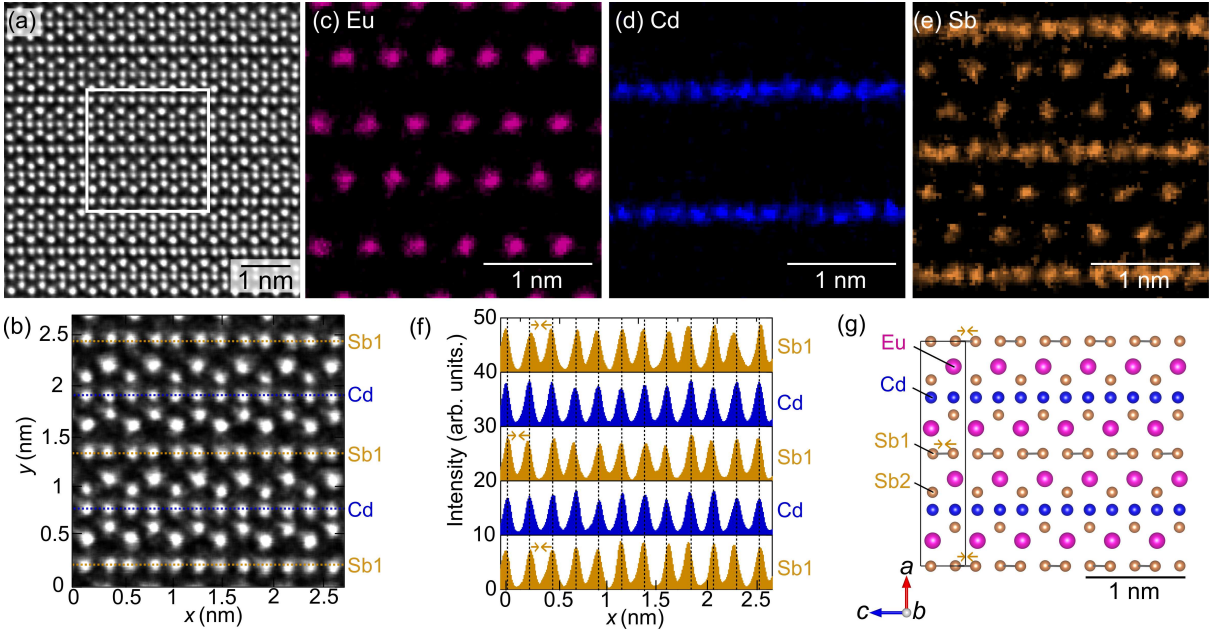}
		\caption{
			(a) Cross-sectional high-angle annular dark-field (HAADF) scanning transmission electron microscopy (STEM) image of the {\ECS} film.
			(b) Higher resolution HAADF-STEM image for the boxed region in (a) and	energy dispersive X-ray spectrometry map for (c) Eu $L$, (d) Cd $L$, and (e) Sb $L$ edges.
			The Eu, Cd, and Sb atoms are aligned as expected for the $AB$Sb$_2$ compounds shown in Fig.~1(a).
			(f) Intensity line profiles of Cd and Sb1 layers extracted from dotted horizontal lines in (b). 
			In contrast to the equally spaced Cd atoms, alternate left and right shifts are clearly resolved in the Sb1 layers, as marked with arrows.
			(g) Cross-sectional $Pnma$ structure, viewed along the $b$-axis.
			Arrows represent the atomic shift or distorted square nets in the Sb1 layers.
		}
		\label{fig3}
	\end{center}
\end{figure}
\begin{figure}
	\begin{center}
		\includegraphics*[bb=0 0 488 396,width=16cm]{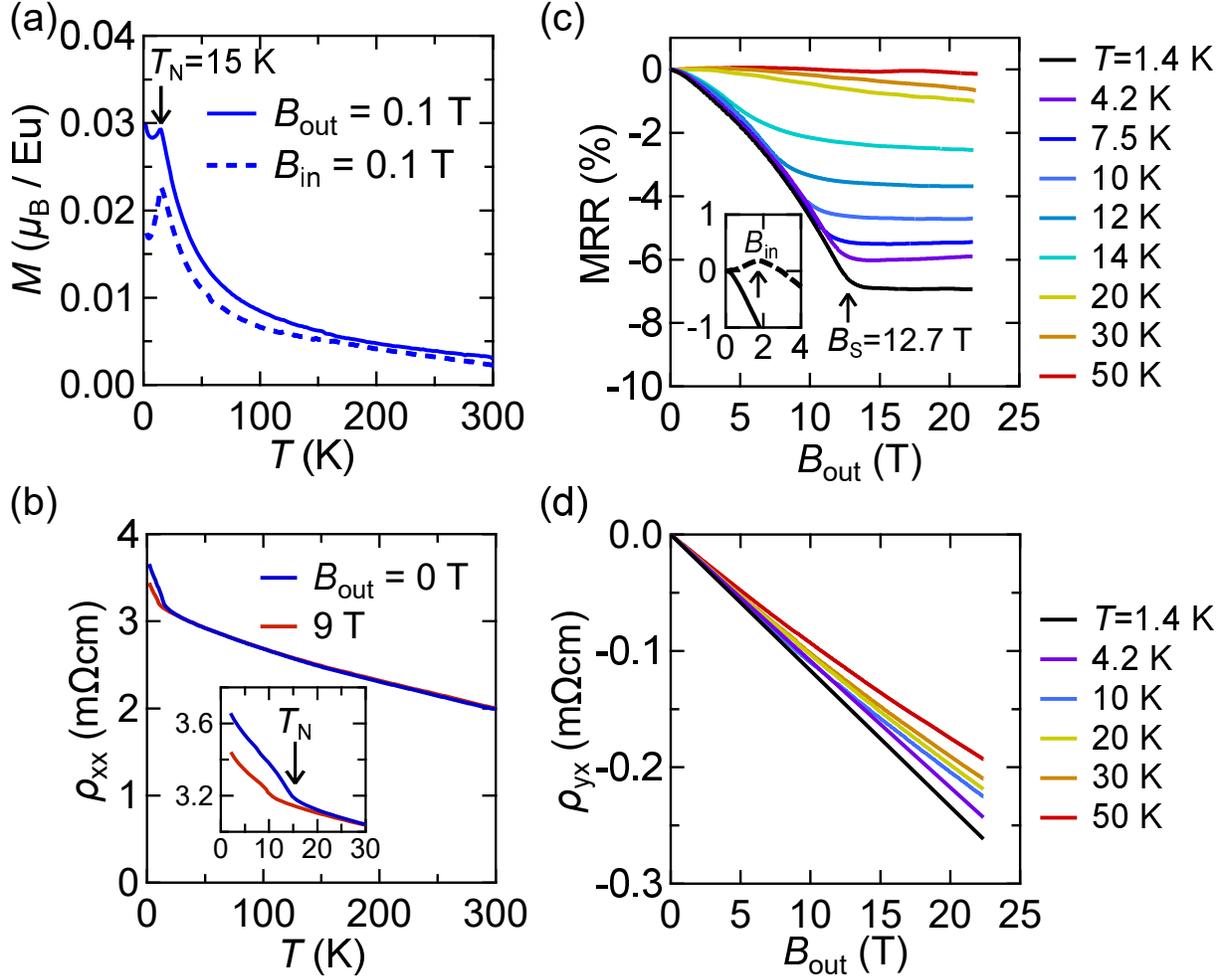}
		\caption{
			(a) Temperature dependence of magnetization $M$, taken with applying the out-of-plane or in-plane magnetic field of 0.1~T.
			Both curves exhibit a kink at ${T_{\mathrm{N}}}=15$~K.
			(b) Temperature dependence of longitudinal resistivity ${\rho_{\mathrm{xx}}}$.
			The inset shows an enlarged view at low temperatures around $T_{\mathrm{N}}$.
			A drastic upturn observed below $15$~K shifts to lower temperatures for $9$~T.
			(c) Magnetoresistance ratios (MRR $=(R_{\mathrm{xx}}(B)-R_{\mathrm{xx}}(0$~$\mathrm{T}))/R_{\mathrm{xx}}(0$~$\mathrm{T})$)
			taken with sweeping the out-of-plane field at various temperatures.
			The saturation field is estimated to be $B_\mathrm{s}=12.7$~T at $1.4$~K.
			The inset compares low-field MRR, taken for the out-of-plane (solid line) and in-plane (dashed line, $I \parallel B$) fields at 1.4~K.
			For the in-plane field, transition from positive to negative magnetoresistance is found at $1.7$~T.
			(d) Hall resistivity $\rho_{\mathrm{yx}}$ taken at various temperatures.
		}
		\label{fig4}
	\end{center}
\end{figure}
\end{document}